\documentclass[11pt,a4paper]{amsart}

\pdfoutput=1

\usepackage{geometry}                
\usepackage{graphicx}
\usepackage{amssymb}
\usepackage{epstopdf}
\usepackage{color}
\usepackage{float}
\usepackage{caption}
\usepackage[table]{xcolor}

\usepackage{url}

\RequirePackage[OT1]{fontenc}
\RequirePackage{amsthm,amsmath}
\usepackage{amsmath, amssymb, amsfonts} 
\usepackage{booktabs}  
\usepackage{graphicx} 

\theoremstyle{plain}

\theoremstyle{definition}

\theoremstyle{remark}

\begin{document}

\title[Identifying mediating variables with graphical models]{Identifying mediating variables with graphical models: an application to the study of causal pathways in people living with HIV}

\author{Adrian Dobra}
\address{Department of Statistics, University of Washington, Seattle, WA, USA}

\author{Katherine Buhikire and Joachim G. Voss}
\address{Frances Payne Bolton School of Nursing, Case Western Reserve University, Cleveland, OH, USA}

\maketitle

\date{\today} 

\begin{abstract}
We empirically demonstrate that graphical models can be a valuable tool in the identification of mediating variables in causal pathways. We make use of graphical models to elucidate the causal pathway through which the treatment influences the levels of fatigue and weakness in people living with HIV (PLHIV) based on a secondary analysis of a categorical dataset collected in a behavioral clinical trial: is weakness a mediator for the treatment and fatigue, or is fatigue a mediator for the treatment and weakness? Causal mediation analysis could not offer any definite answers to these questions.
\end{abstract}

\tableofcontents

\section{Introduction}

Mediating variables transmit the effect of a variable on another variable, and are key in many scientific fields such as social sciences, public health, medicine and psychology for assessing various causal pathways \cite{baron-kenny-1986,pearl-2000,mackinnon-et-2007,fritz-et-2008}. However, in studies that involve several outcome variables that are potentially highly correlated with each other, it is unclear which of these outcome variables are directly influenced by a treatment variable, and which outcome variables are  indirectly influenced by the treatment variable through one or several mediating variables. In these situations, current statistical methods for causal mediation analysis  variables might not be helpful in determining the variables with a potential mediator role.

Our data come from a study of the longitudinal progression of fatigue and weakness in people living with HIV (PLHIV). Fatigue and weakness are closely related when self-rated by PLHIV, and prevalence rates often exceed 50\% for both symptoms regardless of HIV status, treatment status, or age \cite{voss-et-2007, willard-et-2009}.  Nearly forty years into the HIV epidemic, fatigue continues to be the most frequently reported and debilitating symptom for PLHIV \cite{duracinsky-et-2012, jong-et-2010, DiBonaventura-et-2012}, and has been documented extensively with prevalence data ranging from 30-98\% among multiple samples of PLHIV \cite{willard-et-2009,jong-et-2010,al-dakkak-2012,barroso-voss-2013}. Jong et al. \cite{jong-et-2010} identified several predictors of fatigue, including unemployment and inadequate income; combined antiretroviral therapy; and psychological factors such as anxiety and depression. Interestingly, laboratory values did not predict fatigue.  Fatigue has been reported as the most frequent side effect of antiretroviral medications and is associated with worse health status, decreased work productivity, and increased health resource use \cite{DiBonaventura-et-2012}. It interferes with antiretroviral adherence \cite{al-dakkak-2012}, and is one of several factors that predict virologic failure independent of medication adherence measures \cite{marconi-et-2013}. 

Fatigue has been classified into normal daily fatigue, a state due to natural variations in a person’s energy expenditure and circadian rhythms \cite{buseh-et-2008}. Normal fatigue is resolved within a 6\--8h rest and sleep period \cite{pence-et-2009}.  Chronic fatigue on the other hand is hallmarked by persistent or intermittent sensations of tiredness and weariness that last longer than six months and are very unpredictable in duration and intensity \cite{barroso-et-2010}. Rest and sleep do not resolve chronic fatigue but can rather lead to secondary fatigue resulting from de-conditioning and altered sleep patterns \cite{leserman-et-2008,onen-et-2009, jason-et-2010}.

The causal explanations of the origins of fatigue have been divided into three main areas: 1) centrally related immune factors that change brain functions and impact cognition, mood and behavior changes \cite{powers-2017}; 2) neuroendocrine and immune contributors \cite{jason-choi-2008} \--- stressors to the hypothalamic-pituitary-adrenal axis, the autonomic nervous system and the immune system \--- that impact how the body to responds appropriately to stress and inflammation; and 3) peripheral factors such as mitochondrial dysfunction related to fatigue decline in skeletal muscle function – and muscle loss \cite{hidalgo-et-2002}. HIV infection, deconditioning, chronic inflammation and physiological and psychological side effects from antiretroviral treatments such as poor sleep and depression, stigma and anxiety most likely impact all three mechanisms \cite{nixon-et-2005,salahuddin-et-2009}. Muscle weakness in PLHIV is usually a sign of deconditioning and frailty hallmarked by significant involuntary weight loss, increasing slowness, and reduction in grip strength, or muscle atrophy \--- which means the active loss of muscle tissue, or muscle inflammation \--- attack of immune cells of the muscle fibers due to autoimmune or toxicity-related (AZT) processes, or a combination of these factors \cite{benamar-et-2010,dalakas-2009, estanislao-2004,terzian-et-2009, cook-et-2011, chen-et-2012, jenkin-et-2006, sliep-et-2001}.

Episodic or persistent tiredness and exhaustion related to fatigue and weakness are quite common, debilitating, and have major impacts on PLHIV’s social, role, and emotional functioning \cite{keyser-2010,finsterer-mahjoub-2014}. Patients describe the fatigue and weakness sensations with phrases such as ``I feel the same when I wake up than when I went to sleep"; or ``I‘ve got to drag myself" \cite{leavitt-deluca-2010}. The perception of fatigue and weakness is for many patients very difficult to distinguish, however, when asked to rate them, they will distinguish both symptoms. High rates of fatigue and weakness were previously related to the HIV disease or related to side effects from antiretroviral therapies \cite{wantland-et-2011}. Most investigators have not applied adequate longitudinal analytic methodologies to analyze these symptoms together to establish whether a symptom cluster of fatigue and weakness is maintained over time in the same individuals. While fatigue and weakness may have similar sensations and therefore are used interchangeably to describe them, they have very specific recommendations to alleviate them. Practitioners strive in their conversations with patients to better understand whether the patient is indeed suffering from fatigue and/or from weakness. That differentiation would allow clinicians to make better recommendations to alleviate one and/or the other symptom.  

This paper contributes to a better understanding of the relationship between fatigue and weakness in PLHIV based on a secondary analysis of a categorical dataset collected in a behavioral clinical trial. We aim to elucidate the causal pathway through which the treatment influences the levels of fatigue and weakness in PLHIV: is weakness a mediator for the treatment and fatigue, or is fatigue a mediator for the treatment and weakness? We propose a new method for identifying mediating variables, and empirically demonstrate the value of this method to answer these questions.
 
\section{Data description} \label{sec:data}

The data used in our study consists of responses from 609 study participants from a three-month, longitudinal, randomized, controlled trial of HIV-positive individuals. The original study was reviewed by the Committee on the Protection of Human Subjects at the University of California, San Francisco, which accepted the protocol \cite{terzian-et-2009,wantland-et-2008,wantland-et-2011}. There are four variables of interest: (1) signs and symptom checklist fatigue (SSC-F) with four levels (0= absent, 1=mild, 2=moderate, 3=severe); (2) signs and symptom checklist weakness (SSC-W) with four levels (0 = absent, 1=mild, 2=moderate, 3=severe); (3) treatment (IC) with two levels (0= control group, 1 = intervention group); and (4) data collection time (TIME) with three levels (0 = start of the study, 1 = one month, 2 = three months). The fatigue and weakness levels were self scored by the study participants. Out of the 609 individuals, 173 (28.41\%) had only one symptom report, 138 (22.66\%) had two symptom reports, and 298 (48.93\%) had three symptoms recorded. After the removal of the records with missing values, we ended up with 1343 observations. Each observation represents the measurements of weakness and fatigue of one individual at one time point.

In the control group (IC=0), 53 study participants did not complain of either fatigue or weakness at the start of the study (TIME=0), while 208 study participants experienced symptoms consistent with fatigue or weakness or both. At one month (TIME = 1), the corresponding number of study participants decreased to 45 and 143, while at three months (TIME = 2), these numbers were 47 and 135, respectively. In the intervention group (IC=1), 83 study participants did not complain of either fatigue or weakness at the start of the study (TIME=0), while 232 study participants complained of fatigue or weakness or both. At one month (TIME=1), the corresponding numbers of study participants were 69 and 136, while at three months (TIME=2), these numbers were 87 and 105, respectively. With respect to gender, the sample was 195 (32.2\%)  male; 397 (65.19\%)  female; 17 (2.79\%) transgender. With respect to race, the sample was 10 (1.64\%) Asian/Pacific Islander ; 214 (35.14\%) African American/black; 191 (32.36\%) Hispanic/Latino; 8 (1.31\%) Native American Indian; 162 (26.60\%) white/anglo (non-Hispanic); and 24 (3.94\%) other. The mean age of the study participants was 44.15 years (SD = 9.16), with a range between 20-70 years. 

The 1343 observations are cross-classified with respect to SSC-F, SSC-W, TIME and IC in a four-dimensional contingency table with $4 \times 4 \times 3 \times 2 = 96$ cells \--- see Table \ref{tab:fourwaytable}. In this contingency table, a cell count represents the number of observations at one time point associated with the control or with the intervention group who had a certain weakness and a certain fatigue level.
 
 \begin{table}
\caption{Data recorded in the behavioral clinical trial. Study participants are cross-classified by fatigue (SSC-F), weakness (SSC-W),  time point (TIME) and treatment group (IC).}
{\begin{tabular}{ccccccccccc}\toprule
   && IC  & \multicolumn{4}{c}{0} & \multicolumn{4}{c}{1}\\
  SSC-F & TIME & SSC-W & 0 & 1 & 2 & 3 & 0 & 1 & 2 & 3\\ 
  \midrule
0 & 0 &&  53 & 10 & 6 & 6 &  83 & 11 & 6 & 5\\
1 & 0 &&  10 & 35 & 7 & 0 &  15 & 39 & 13 & 1\\
2 & 0 &&   8 & 18 & 46 & 12 &  16 & 28 & 41 & 4\\ 
3 & 0 &&   3  & 8 & 17 & 22 &  4 & 6 & 19 & 24\\ 
0 & 1 &&  45 & 5 & 5 & 4 &  69 & 9 & 6 & 3\\
1 & 1 &&  10 & 22 & 6 & 0 &  11 & 18 & 5 & 0\\
2 & 1 &&  10 & 14 & 37 & 4 &  10 & 14 & 22 & 3\\
3 & 1 &&   4 & 1 & 4 & 17 &   6 & 4 & 13 & 12\\
0 & 2 &&  47 & 9 & 2 & 5 &  87 & 5 & 7 & 2\\ 
1 & 2 &&  13 & 24 & 4 & 1 &  11 & 12 & 6 & 2\\ 
2 & 2 &&  15 & 12 & 22 & 4 &  12 & 12 & 4 & 1\\ 
3 & 2 &&   2 & 0 & 5 & 17  &   3 & 3 & 3 & 12\\  \bottomrule
\end{tabular}}
\label{tab:fourwaytable}
\end{table}

\section{Approach} \label{sec:approach}

Hierarchical loglinear models \cite{bishop-et-1975,agresti-1990} give statistical representations of joint distributions between two or more categorical variables without assuming that some of the variables involved are outcomes, while the remaining variables are independent. The main effects associated with every variable are present in each loglinear model. Two loglinear models involving the same variables may differ in terms of the interaction terms involving two, three or more variables they include. The interaction terms are included in a loglinear model in a hierarchical manner: the presence of any higher order interaction term requires the presence of any of the lower order interactions terms which involve a subset of the variables present in the higher order term. The selection of the interaction structure in hierarchical loglinear models is computationally very difficult due to the exponential increase in the number of possible hierarchical loglinear models: while there are nine models with $3$ variables and 7580 models with $5$ variables, there are about $5.6\times 10^{22}$ models with $8$ variables \cite{dellaportasforster1999}. Selection of hierarchical loglinear models has been widely discussed in the statistical literature \cite{fienberg-1970,edwardshavranek1985,agresti-1990,whittaker1990}. More recent approaches that work well for high-dimensional sparse contingency tables  involve Bayesian Markov chain Monte Carlo (MCMC) algorithms\cite{madiganraftery1994,madiganyork1995,madiganyork1997,dellaportasforster1999,tarantola2004,dellaportastarantola2005,dobra-massam-2010,dobra-lenkoski-2011,dobra-mohammadi-2018}.  

A loglinear model is hierarchical if and only if the presence of a higher-order interaction term requires the presence of any or all of its lower-order interaction terms \cite{bishop-et-1975}. The model may not be identifiable without imposing some constraints on the interaction terms \cite{agresti-1990}.  Some hierarchical loglinear models are also graphical. The distinction between a graphical loglinear model and a hierarchical loglinear model that is not graphical can be easily understood by constructing an undirected graph $G$, with vertices $\mathcal{B}$ and edges $E$, that is associated to a hierarchical loglinear model $\mathcal{M}$ \cite{whittaker1990} \--- see Figure \ref{fig:1} for an example. Specifically, an edge $e=(b_{1},b_{2})$ appears in $G$ if and only if the observed variables $X_{b_{1}}$ and $X_{b_{2}}$ appear together in an interaction term of $\mathcal{M}$. The graph $G$ is called the interaction graph of model $\mathcal{M}$ \cite{lauritzen-1996}. Then $\mathcal{M}$ is graphical if and only if the subsets of $\mathcal{B}$ that are the vertices of the complete subgraphs of $G$ that are maximal with respect to inclusion, are also maximal interaction terms in $\mathcal{M}$ \cite{whittaker1990,lauritzen-1996}.  If $\mathcal{M}$ is graphical, the absence of an edge $e=(b_{1},b_{2})$ in $G$ is  equivalent with the conditional independence of variables $X_{b_{1}}$ and $X_{b_{2}}$ given the rest of the variables under the joint distribution for $X_{\mathcal{B}}$.  Moreover, if there is no path in $G$ from vertex $b_{1}$ to vertex $b_{2}$, then $b_{1}$ and $b_{2}$ are in two distinct and fully connected components of $G$, and the corresponding variables $X_{b_{1}}$ and $X_{b_{2}}$ are independent. Most importantly, for any three disjoint sets $A$, $B$ and $C$ such that $\mathcal{B}= A\cup B\cup C$, we say that $B$ separates $A$ and $C$ in $\mathcal{G}$ if any path that connects a vertex in $A$ with a vertex in $C$ contains at least one vertex in $B$. If $B$ is a complete subset of $G$ (that is, if the subgraph of $G$ induced by $B$ does not have any missing edge), we say that $B$ is a separator of $G$, and that $(A,B,C)$ is a weak decomposition of $G$ \cite{lauritzen-1996}. If $(A,B,C)$ is a weak decomposition of $G$, then $X_A$ and $X_B$ are conditional independent given $X_C$. 

We propose the following method for the determination of mediating variables:

\begin{enumerate}
 \item[] Step 1. Determine a graphical model that is best supported by the data.
 \item[] Step 2. Determine a weak decomposition $(A,B,C)$ of the undirected graph associated with the graphical model determined in the first step such that the treatment variable is the only element of set $A$. Then the variables that belong to set $B$ are candidate mediator variables.
 \item[] Step 3. Establish mediation of the variables in $B$ using criteria from causal mediation analysis.
\end{enumerate}

The last step is needed because the interaction graphs we defined above are undirected, therefore they do not convey any structural information about the directionality of the relationships represented by the edges of these graphs. This distinction is key in causal inference \cite{pearl-2000}.

It is possible that the hierarchical loglinear model that is best supported by the data is not graphical. This is more likely to happen when the number of observed variables becomes large. In these cases, we recommend explicitly restricting the search to the class of graphical hierarchical models. Such a restriction is beneficial from a computational point of view because fewer possible models will be considered. We remark that our proposed method for the identification of mediating variables is not restricted to categorical data, and can be used for datasets that involve continuous variables, or continuous and discrete variables. The existing statistical literature has defined rich families of graphical models that involve only continuous variables (e.g., Gaussian graphical models) or variables of mixed type \--- see, for example, \cite{jordan-2004,koller-friedman-2009}. 

\section{Application} \label{sec:application}

We follow our proposed three step procedure to determine the mediator roles of weakness and fatigue for the categorical data described in Section \ref{sec:data}. The identification of the most relevant interactions among SSC-F, SSC-W, TIME and IC is essential due to the ability of hierarchical loglinear models to capture multivariate association structures that show how weakness and fatigue evolve over time, as well as their change across treatment groups. 

\subsection{Loglinear models for understanding the relationship between weakness and fatigue over time} \label{sec:wft}

We begin by exploring the relationship between weakness (SSC-W) and fatigue (SSC-F) adjusted for data collection time (TIME) without adjusting for treatment (IC). To this end, we examine all possible hierarchical loglinear models that involve SSC-W, SSC-F and TIME \--- see Table \ref{tab:2}.
 
\begin{table}
\caption{Hierarchical loglinear models for SSC-W, SSC-F and TIME. Here $G^2$ stands for the log-likelihood ratio test statistic, and DF stands for degrees of freedom.}
{\begin{tabular}{ccccc}
\toprule
Model & Minimal sufficient statistics & $G^2$ & DF & p-value\\ \midrule
1 & [SSC-W][SSC-F][TIME]	& 859.80	& 39	& 0\\ 
 2 & [SSC-W][SSC-F,TIME]	& 	841.50	& 33	& 0\\ 
3 & [SSC-W,TIME][SSC-F]	& 	826.73	& 33	& 0\\ 
4 & [SSC-W,SSC-F][TIME]	& 51.86	& 30	& 0.010\\ 
5 & [SSC-W,SSC-F][SSC-W,TIME]	& 18.79	& 24	& 0.763\\ 
6 & [SSC-W,SSC-F][SSC-F,TIME]	&	33.56	& 24	& 0.093\\
7 & [SSC-W,TIME][SSC-F,TIME]	& 808.44	& 27	& 0\\
8 & [SSC-W,SSC-F][SSC-W,TIME][SSC-F,TIME] 	& 14.75	& 18	& 0.679\\ \bottomrule
\end{tabular}}
\label{tab:2}
\end{table}

Only Models 5, 6 and 8 from Table \ref{tab:2} have a p-value associated with the log-likelihood ratio test statistic $G^2$ greater that $0.05$, and fit the data well \cite{bishop-et-1975}. These three loglinear models contain the interaction between SSC-W and SSC-F which implies that  these data provide evidence of a direct relationship between weakness  and fatigue. We determine which of these three models is best supported by the data by performing two pairwise comparisons \--- see Table \ref{tab:3}. First, we compare Models 6 and 8 by testing the null hypothesis that there is no interaction between SSC-W and TIME, and found Model 8 to be superior to Model 6 (p-value = 0.004). We subsequently compare Models 5 and 8 by testing the null hypothesis of no interaction between SSC-F and TIME. In this case the difference between the values of $G^2$ of these two models was 4.04, hence we fail to reject the null hypothesis of no interaction between SSC-F and TIME (p-value = 0.671). Thus Model 5 is best supported by the data. This model suggests that TIME does not have a direct effect on fatigue, but it has a direct effect on weakness. Model 5 also indicates that the effect of TIME on fatigue is indirect. 

\begin{table}
\caption{Comparison between the three hierarchical loglinear models for SSC-W, SSC-F and TIME that give a good fit to the data.}
{\begin{tabular}{cccc}
\toprule
Comparison &	$G^2$ difference	& DF difference	& p-value\\ \midrule
Model 6 vs. Model 8 & 18.82	& 6	& 0.004\\ 
Model 5 vs. Model 8 & 4.04	& 6	& 0.671\\ \bottomrule
\end{tabular}}
\label{tab:3}
\end{table}

 \subsection{Loglinear models for understanding the relationship between weakness and fatigue over time as influenced by the treatment} \label{sec:wfti}

We investigate whether the multivariate relationships among SSC-W, SSC-F and TIME still hold when controlling for the treatment (IC). For this purpose, we examine the fit of the three hierarchical loglinear models that include the interactions present in Model 5 (see Table \ref{tab:3}) together with interactions between weakness and treatment group (Model 9), fatigue and treatment group (Model 10), or both (Model 11). The results are summarized in Table \ref{tab:4}. 

\begin{table}
\caption{Hierarchical loglinear models for SSC-W, SSC-F, TIME and IC.}
{\begin{tabular}{ccccc}
\toprule
Model & Minimal sufficient statistics & $G^2$& DF & p-value\\ \midrule
9	& [SSC-W,SSC-F][SSC-W,TIME][SSC-W,IC] & 	67.11 & 68 & 0.507\\
 10 & [SSC-W,SSC-F][SSC-W,TIME][SSC-F,IC] & 70.64 & 68 & 0.389\\ 
 11 & [SSC-W,SSC-F][SSC-W,TIME][SSC-W,IC][SSC-F,IC] & 58.45 & 65 & 0.704\\ \bottomrule
\end{tabular}}
\label{tab:4}
\end{table}

Model 5 is nested in each of Models 9, 10 and 11, and these three models fit the data well (p-value $>$ 0.05). Thus the direct relationships between SSC-W and SSC-F, and between SSC-W and TIME still hold in the presence of the intervention. A comparison of Models 10 and 11 based on their $G^2$ goodness-of-fit statistics leads to rejection of the null hypothesis of no interaction between SSC-F and IC (p-value = 0.007) \--- see Table \ref{tab:5}. A similar comparison of Models 9 and 11 leads us to fail to reject the null hypothesis of no interaction between SSC-W and IC (p-value = 0.034). As such, the loglinear model with the best fit among the three models we considered is Model 9. This model shows that SSC-W and SSC-F are directly related when in the presence of IC and TIME, and that there is no direct effect of TIME and IC on fatigue. However, both TIME and IC have an indirect effect on fatigue that is mediated by weakness. 
 
\begin{table}
\caption{Loglinear model selection for SSC-W, SSC-F, TIME and IC.}
{\begin{tabular}{cccc}
\toprule
Comparison &	$G^2$ difference & DF difference	& p-value\\ \midrule
Model 9 vs. Model 11 & 8.66	& 3	& 0.034\\
Model 10 vs. Model 11 & 12.19 & 3	& 0.007\\  \bottomrule
\end{tabular}}
\label{tab:5}
\end{table}

\begin{figure}
\centering
\includegraphics[scale=1]{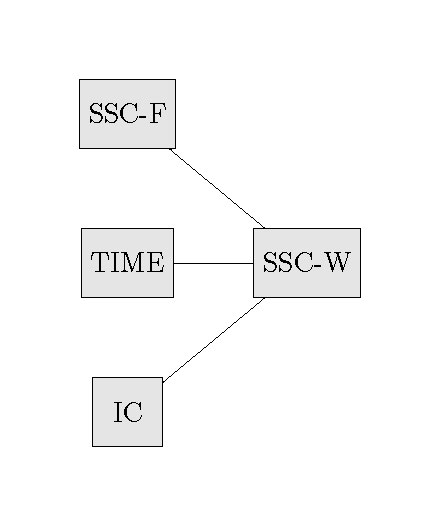}
\caption{Interaction graph associated with Model 9. This is the graphical loglinear model with minimal sufficient statistics [SSC-W,SSC-F][SSC-W,TIME][SSC-W,IC]. Each vertex of this graph corresponds with an observed variable. Each edge of this graph corresponds with a pairwise interaction term of Model 9.}
\label{fig:1}
\end{figure}

The interaction graph associated with Model 9 is shown in Figure \ref{fig:1}. The four variables SSC-W, SSC-F, TIME and IC are each represented as vertices of this graph. Two vertices are linked by an edge if Model 9 includes an interaction term between the corresponding variables. Since Model 9 involves three pairwise interaction terms, the graph in Figure \ref{fig:1} has three edges. This model is graphical since each edge corresponds with a maximal interaction term of Model 9, namely, the three pairwise interaction terms [SSC-W,SSC-F], [SSC-W,TIME] and [SSC-W,IC]. Since each path that connects SSC-F and TIME, or SSC-F and IC, or TIME and IC must go through SSC-W, Model 9 is the model of conditional independence of SSC-F, TIME and IC given SSC-W.
Furthermore, each path that connects vertex IC with vertices SSC-F or TIME must go through SSC-W, thus $(\{\mbox{IC}\},\{\mbox{SSC-W}\},\{\mbox{SSC-F},\mbox{TIME}\})$ is a weak decomposition of the interaction graph in Figure \ref{fig:1}. Thus the interaction structure of Model 9 suggests that SSC-W is a candidate mediator variable, and SSC-F cannot be a mediator variable.

We note that a graphical representation of Model 5 is obtained by eliminating the vertex IC and the edge between SSC-W and IC from the graph in Figure \ref{fig:1}. Model 5 is also a graphical model: it is the model of conditional independence of SSC-F and TIME given SSC-W.

\subsection{Model validation}
 
Our approach for selecting Models 5 and 9 was solely based on the data available. While such a model selection strategy is widely used in modern statistical literature, it potentially raises concerns related to whether the same hierarchical loglinear models would be selected if new samples involving the same variables became available. In order to address this question, we design an objective procedure for assessing the out of sample generalizability of the models we selected. This procedure involves repeatedly sampling a fraction of the available samples, and determining the most relevant loglinear models only based on the selected samples.

Instead of employing once again the hypothesis testing approach that led us to selecting Models 5 and 9, we follow a different approach for model determination. Here the most relevant hierarchical loglinear model is considered to be the loglinear model with the smallest Akaike information criterion (AIC) \cite{hojsgaard-et-2012}. The determination of the loglinear model with the smallest AIC is performed in two different ways:

\begin{itemize}
\item {\it Forward search}: we start at the loglinear model of independence that involves only the main effects. This model is the current model at the first iteration. At subsequent iterations, the current loglinear model is improved by including additional interaction terms that are not in the model such that the model is still hierarchical. The search stops when the inclusion of additional interaction terms does not lead to a hierarchical loglinear model with a smaller AIC.
\item {\it Backward search}: we start at the saturated loglinar model that involves all possible interaction terms. This model is the current model at the first iteration. At subsequent iterations, the current loglinear model is improved by deleting interaction terms from the model such that the model is still hierarchical. The search stops when the deletion of additional interaction terms does not lead to a hierarchical loglinear model with a smaller AIC.
\end{itemize}

The forward and backward search procedures have been implemented with functions from the \texttt{R} \cite{r-manual} package \texttt{gRim} \cite{hojsgaard-et-2012}. We consider that we have successfully identified the model with the smallest AIC if both the forward and the backward search lead to the same hierarchical loglinear model. Based on all 1343 available samples, Model 5 turns out to be the hierarchical loglinear model with the smallest AIC that involves variables SSC-W, SSC-F and TIME. Similarly, based on all the available samples, Model 9 turns out to be the hierarchical loglinear model with the smallest AIC that involves variables SSC-W, SSC-F, TIME and IC. This finding supports our claim that Models 5 and 9 are the most representative hierarchical loglinear models for these data. We note that, in Section \ref{sec:wfti}, we selected Model 9 by considering only three candidate hierarchical loglinear models. This shortcoming is addressed by the combined forward and backward search methods that consider as candidates all possible hierarchical loglinear models with 4 variables.

Does the AIC-based model selection strategy hold when only a fraction of the available samples are used? To answer this question, for each $q=1\%, 2\%,\ldots,99\%$, we repeat the following three steps 10000 times:

\begin{enumerate}
\item Sample without replacement a proportion $q$ of the available samples.
\item Determine the most relevant hierarchical loglinear model that involves variables SSC-W, SSC-F and TIME based on the samples selected at Step 1.
\item Determine the most relevant hierarchical loglinear model that involves variables SSC-W, SSC-F, TIME and IC based on the samples selected at Step 1.
\end{enumerate}

The determination of the most relevant hierarchical loglinear model is performed based on the forward and backward search methods as described above. When the forward and backward search methods do not stop at the same loglinear model, no model is recorded for that particular sampled dataset.

\begin{figure}
\centering
\includegraphics[scale=0.5]{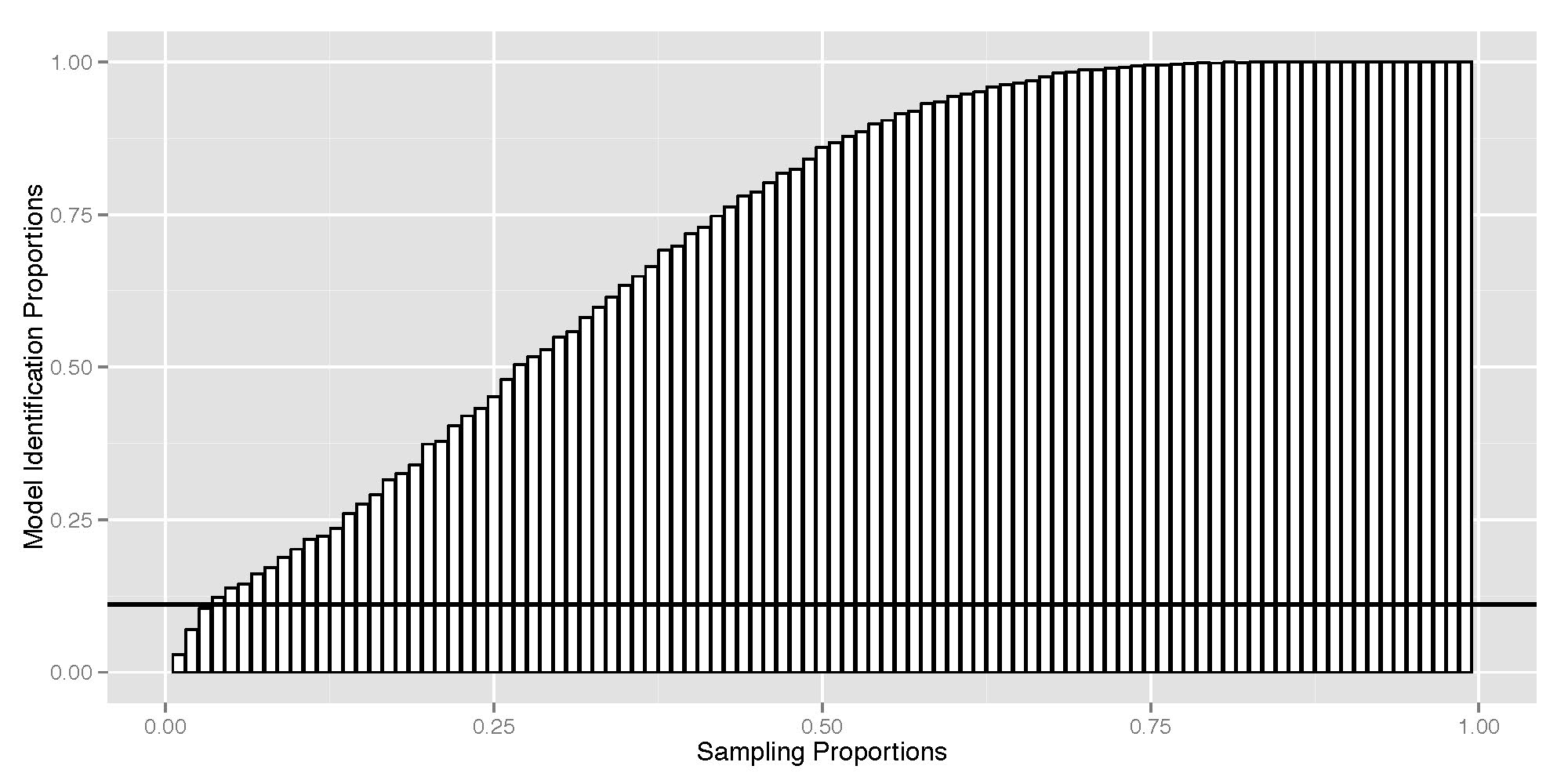}
\caption{Out of sample validation for Model 5. Each bar represents the proportion of times based on 10000 sampled datasets Model 5 has been identified to be the loglinear model with the smallest AIC involving variables SSC-W, SSC-F and TIME. From left to right, the bars correspond with sampling proportions of $1\%,2\%,\ldots,99\%$. The horizontal line shows the empirical probability $1/9=0.111$ of selecting Model 5 at random from the set of nine hierarchical loglinear models with three categorical variables.}
\label{fig:2}
\end{figure}

\begin{figure}
\centering
\includegraphics[scale=0.5]{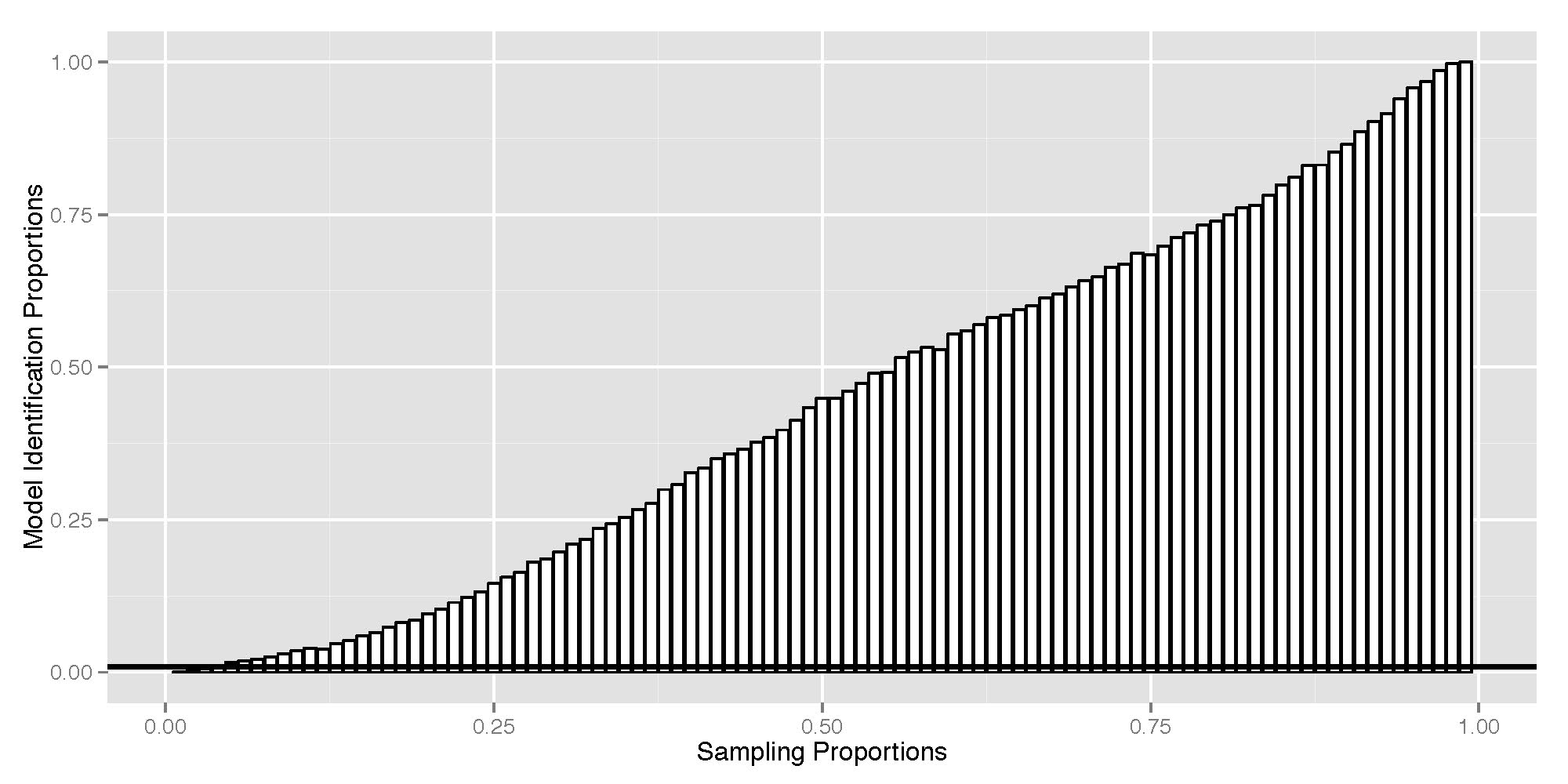}
\caption{Out of sample validation for Model 9. Each bar represents the proportion of times based on 10000 sampled datasets Model 9 has been identified to be the loglinear model with the smallest AIC involving variables SSC-W, SSC-F, TIME and IC. From left to right, the bars correspond with sampling proportions of $1\%,2\%,\ldots,99\%$. The horizontal line shows the empirical probability $1/114=0.0087$ of selecting Model 9 at random from the set of 114 hierarchical loglinear models with four categorical variables.}
\label{fig:3}
\end{figure}

We objectively assess the out of sample properties of our model selection strategy from Sections \ref{sec:wft} and \ref{sec:wfti} by reporting the proportions of times Models 5 and 9 were successfully identified \--- see Figures \ref{fig:2} and \ref{fig:3}. When 50\% or more samples are selected from the available data, Model 5 is determined at Step 2 for at least 80\% of times. When 85\% or more samples are selected, Model 9 is determined at Step 3 for at least 80\% of times. The increased number of samples needed to recover Model 9 relative to the number of samples needed to recover Model 5 is consistent with the size of the set of models they are selected from: while there are only 9 hierarchical loglinear models with 3 variables, there are 114 hierarchical loglinear models with 4 variables. If a hierarchical loglinear model is selected at random from the pool of hierarchical loglinear models with 3 and 4 variables in the absence of any data, Model 5 is chosen with a probability of $1/9=0.111$, and Model 9 is chosen with a probability of $1/114=0.0087$. Figures \ref{fig:2} and \ref{fig:3} show that the probabilities of selecting Model 5 and Model 9 become significantly higher as more samples are employed in the selection process. These findings confirm that Models 5 and 9 are representative for the associations that exist among SSC-W, SSC-F, TIME and IC which provides support for the generalizability of our conclusions. 

\subsection{Causal mediation analysis} \label{sec:causal}

Model 9, the model of conditional independence of SSC-F, TIME and IC given SSC-W posits that: (i) there is no direct effect of the treatment on fatigue; (ii) there is a direct effect of the treatment on weakness; and (iii) there is a direct effect of weakness on fatigue. This suggest that weakness is a mediator variable. Due to the indirect effect of the treatment on fatigue without a direct effect, Model 9 provides evidence for the strongest type of mediation which was called full mediation by Baron and Kenny \cite{baron-kenny-1986}.

In the sequel we want to ascertain whether we can reach the same conclusions by employing a modeling framework for conducting causal mediation analysis. We follow methods and algorithms from Imai et al. \cite{imai-et-2010,imai-et-2013} implemented in the  \texttt{R} \cite{r-manual} package \texttt{mediation} \cite{tingley-et-2014}. Their model-based causal mediation analysis builds on the procedure of Baron and Kenny \cite{baron-kenny-1986}, and comprises the specification of two statistical models. The first model is called the mediator model, and specifies the conditional distribution of the mediator given the treatment and other pre-treatment covariates. The second model is called the outcome model, and specifies the conditional distribution of the outcome, given the treatment, the mediator, and the pre-treatment covariates. After these two models are fit separately, the size of the indirect effect of the treatment on the outcome is estimated by the average causal mediation effects (ACME), while the direct effect of the treatment on the outcome is estimated by the average direct effects (ADE). The sums of the indirect and the direct effects are called the total effects.

First, we estimate the causal effects of the treatment on fatigue with weakness as the mediator variable. The mediator model is an ordinal logistic regression model of SSC-W conditional on IC and TIME. The outcome model is an ordinal logistic regression model of SSC-F conditional on SSC-W, IC and TIME. The results obtained with the \texttt{mediate} function from the \texttt{mediation} R package are shown in Table \ref{tab:fatigue}. With a single exception, the ACME of treatment on fatigue are statistically significant at the $\alpha = 0.05$ level for the treated and the control groups. The ADE of the treatment on fatigue are not statistically significant for both groups. However, total effects of the treatment on fatigue are statistically significant at the $\alpha = 0.05$ level. As such, the indirect effects of the treatment on fatigue mediated by weakness are strong, while there does not seem to be any evidence of direct effects of the treatment on fatigue.

\begin{table}
\caption{Estimated causal effects of the treatment on fatigue with weakness as mediator variable. The table shows nonparametric bootstrap confidence intervals based on 2500 Monte Carlo draws with the percentile method. Here ACME stands for average causal mediation effects, and ADE stands for average direct effects.}
{\begin{tabular}{ccccc}
\toprule
        & Pr(SSC-F=0)  & Pr(SSC-F=1) & Pr(SSC-F=2) & Pr(SSC-F=3)\\
ACME (control) &  0.0605 & 0.006589 & -0.0268 & -0.0403\\
2.5\%            &  0.0294 & 0.000664 & -0.0494 & -0.0633\\
97.5\%          &  0.0983 & 0.014627 & -0.0136 & -0.0174\\
p-value         & $<$0.0001 & 0.028800  & 0.0008  & 0.0008\\ \midrule
ACME (treated)   & 0.0613  & 0.005649 &-0.0275 & -0.0394\\
2.5\%             & 0.0297 & -0.000046 & -0.0507 & -0.0621\\
97.5\%           & 0.0988  & 0.013216 & -0.0139 & -0.0166\\
p-value          & $<$0.0001  & 0.052000 &  0.0008  & 0.0008\\ \midrule
ADE (control)  & 0.00792  & 0.000607 & -0.00267 & -0.00585\\
2.5\%            & -0.02325 & -0.001627 & -0.01370 & -0.02762\\
97.5\%           & 0.03867  & 0.002977  & 0.00749  & 0.01670\\
p-value          & 0.62320  & 0.644000  & 0.62320  & 0.62320\\ \midrule
ADE (treated)  & 0.00866  & -0.000334 & -0.00342 & -0.0049\\
2.5\%            & -0.02544 & -0.003323 & -0.01734 & -0.0229\\
97.5\%           & 0.04271  & 0.001852  & 0.00987  & 0.0137\\
p-value          & 0.62320  & 0.633600 & 0.62320  & 0.6232\\ \midrule
Total Effect    & 0.0692  & 0.006256  & -0.0302 & -0.0452\\
2.5\%            & 0.0269 & 0.000358  & -0.0568 & -0.0748\\
97.5\%          & 0.1196 & 0.013460 & -0.0130 & -0.0154\\
p-value         & 0.0016 & 0.040000 & 0.0032  & 0.0032\\ \bottomrule
\end{tabular}}
\label{tab:fatigue}
\end{table}

Second, we estimate the causal effects of the treatment on weakness with fatigue as the mediator variables. The mediator and the outcome models are similar to the ones we specified before with SSC-W and SSC-F replacing each other. The results are presented in Table \ref{tab:weakness}. In this case, most of the ACME, the ADE and the total effects of the treatment on weakness are statistically significant. This suggests that the treatment has strong direct as well as indirect effects mediated by fatigue on weakness. Baron and Kenny \cite{baron-kenny-1986} refer to the presence of both direct and indirect effects as partial mediation. 

\begin{table}
\caption{Estimated causal effects of the treatment on weakness with fatigue as mediator variable. The table shows nonparametric bootstrap confidence intervals based on 2500 Monte Carlo draws with the percentile method. Here ACME stands for average causal mediation effects, and ADE stands for average direct effects.}
{\begin{tabular}{ccccc}
\toprule
        & Pr(SSC-W=0)  & Pr(SSC-W=1) & Pr(SSC-W=2) & Pr(SSC-W=3)\\
ACME  (control)  & 0.0539  & 0.00148  & -0.02766 & -0.02774\\
2.5\%       &       0.0132 & -0.00603 & -0.04233 & -0.04571\\
97.5\%     &        0.0841  & 0.00893 & -0.00635 & -0.00599\\
p-value    &       0.0048  & 0.78160  & 0.00400  & 0.01120\\ \midrule
ACME  (treated)   & 0.0570 & -0.00537 & -0.02933 & -0.02231\\
2.5\%        &      0.0140  & -0.01313 & -0.04507 & -0.03718\\
97.5\%      &      0.0887  & 0.00348 & -0.00749 & -0.00449\\
p-value    &      0.0040  & 0.25120  & 0.00400  & 0.01200\\ \midrule
ADE (control)  &   0.0566  & 0.000629 & -0.0245 & -0.0327\\
2.5\%     &         0.0253 & -0.004313 & -0.0396 & -0.0531\\
97.5\%   &          0.0915  & 0.004686 & -0.0101 & -0.0148\\
p-value  &         $<$0.0001  & 0.891200  & $<$0.0001  & $<$0.0001\\ \midrule
ADE (treated)  &  0.0597 & -0.00622 & -0.0262 & -0.0273\\
2.5\%       &      0.0264 & -0.01240 & -0.0421 & -0.0450\\
97.5\%      &      0.0961 & -0.00123 & -0.0112 & -0.0127\\
p-value   &        $<$0.0001 &   0.00560  & $<$0.0001  & $<$0.0001\\ \midrule
Total Effect  &   0.1136 & -0.00474 & -0.0538 & -0.0550\\
2.5\%     &       0.0621 & -0.01327 & -0.0743 & -0.0802\\
97.5\%   &        0.1593  & 0.00361 & -0.0283 & -0.0299\\
p-value   &      $<$0.0001 &  0.27040 & $<$0.0001 &  $<$0.0001\\ \bottomrule
\end{tabular}}
\label{tab:weakness}
\end{table}

\section{Discussion}

Since fatigue and weakness may have similar sensations, PLHIV experience them jointly, and consequently the measurements of these two symptoms are highly correlated. As such, there is no surprise that a direct effect between fatigue and weakness was identified when employing graphical loglinear models and in the causal mediation analysis. In that regard, the two modeling frameworks led to the same expected conclusion. However, our causal mediation analysis did not offer any definite answer to the question of whether fatigue is a mediator variable for the effects of the treatment on weakness, or whether weakness is a mediator variable for the effects of treatment on fatigue. This is an important question because, in order to maximize the benefits of  interventions that deal with fatigue and weakness, the causal pathways through which the proposed treatments affect the two symptoms need to be established.

Our causal mediation analysis from Section \ref{sec:causal} found evidence of partial mediation of fatigue, and also evidence of full mediation of weakness. Baron and Kenny \cite{baron-kenny-1986} argue that mediation is stronger when no direct effect of the treatment on the outcome is found, but there is evidence of an indirect effect. According to this argument, we could conclude that weakness is more likely to be a mediator than fatigue for this particular treatment. Nevertheless, the literature on mediation analysis points out flaws in Barron and Kenny's criteria for establishing the strength of mediation. For example, Zhao et al. \cite{zhao-et-2010} argues that the presence of a direct effect can represent evidence for the presence of other unobserved mediators. Consequently, the absence of a direct effect should not constitute a measure of the strength of mediation. Instead, Zhao et al. \cite{zhao-et-2010}  propose using the size of the indirect effect of the treatment on the outcome as a more appropriate measure of mediation strength. However, in this particular application, the estimated ACME from Tables \ref{tab:fatigue} and \ref{tab:weakness} do not seem to provide decisive evidence about when the indirect effects are stronger: when weakness is the mediator, or when fatigue is the mediator?

For these reasons causal mediation analysis was not helpful in differentiating between weakness and fatigue as mediator variables. On the other hand, our proposed approach which is based on learning the graphical model that is best supported by the data provides a clearer answer to this question. The model of conditional independence of fatigue, treatment and time given weakness we determined in Section \ref{sec:approach} shows that weakness is a mediator variable, and that fatigue is not a mediator variable.

\section*{Acknowledgement(s)}

The authors would like to thank the members of the International Nursing Network for HIV/AIDS Research for providing access to data used in this paper.

\section*{Funding}

The work of AD was partially supported by the National Science Foundation Grant DMS/MPS-1737746 to University of Washington. 


\begin{thebibliography}{10}
\providecommand{\MR}{\relax\unskip\space MR }
\providecommand{\url}[1]{\normalfont{#1}}
\providecommand{\urlprefix}{Available at }

\bibitem{agresti-1990}
A. Agresti, \emph{Categorical Data Analysis}, John Wiley \& Sons, New York,
  1990.

\bibitem{al-dakkak-2012}
I. Al-{D}akkak, S. Patel, E. Mc{C}ann, A. Gadkari, G. Prajapati, and E.M.
  Maiese, \emph{The impact of specific {H}{I}{V} treatment-related adverse
  events on adherence to antiretroviral therapy: a systematic review and
  meta-analysis}, AIDS Care 25 (2012), pp. 400--414.

\bibitem{baron-kenny-1986}
R.M. Baron and D.A. Kenny, \emph{The moderator-mediator variable distinction
  in social psychological research: Conceptual, strategic, and statistical
  considerations}, Journal of Personality and Social Psychology 51 (1986), pp.
  1173--1182.

\bibitem{barroso-et-2010}
J. Barroso, B.G. Hammill, J. Leserman, N. Salahuddin, J.L. Harmon, and B.W.
  Pence, \emph{Physiological and psychosocial factors that predict
  {H}{I}{V}-related fatigue}, AIDS and behavior 14 (2010), pp. 1415--1427.

\bibitem{barroso-voss-2013}
J. Barroso and J. Voss, \emph{Fatigue in {H}{I}{V} and {A}{I}{D}{S}: An
  analysis of evidence}, Journal of the Association of Nurses in AIDS Care 24
  (2013), pp. S5--14.

\bibitem{benamar-et-2010}
K. Benamar, S. Addou, M. Yondorf, E.B. Geller, T.K. Eisenstein, and M.W. Adler,
  \emph{Intrahypothalamic injection of the {H}{I}{V}-1 envelope glycoprotein
  induces fever via interaction with the chemokine system}, The Journal of
  Pharmacology and Experimental Therapeutics 332 (2010), pp. 549--553.

\bibitem{bishop-et-1975}
Y.M.M. Bishop, S.E. Fienberg, and P.W. Holland, \emph{Discrete {M}ultivariate
  {A}nalysis: {T}heory and {P}ractice}, M.I.T. Press, Cambridge, MA, 1975.

\bibitem{buseh-et-2008}
A. Buseh, S.T. Kelber, P.E. Stevens, and C.G. Park, \emph{Relationship of
  symptoms, perceived health, and stigma with quality of life among urban
  {H}{}{V}-infected {A}frican {A}merican men}, Public Health Nursing 25 (2008),
  pp. 409--419.

\bibitem{chen-et-2012}
W.T. Chen, S.Y. Lee, C.S. Shiu, J.M. Simoni, C. Pan, M. Bao, and H. Lu,
  \emph{Fatigue and sleep disturbance in {H}{I}{V}-positive women: a
  qualitative and biomedical approach}, Journal of Clinical Nursing 22 (2012),
  pp. 1262--1269.

\bibitem{cook-et-2011}
P.F. Cook, K.H. Sousa, E.E. Matthews, P.M. Meek, and J. Kwong, \emph{Patterns
  of change in symptom clusters with {H}{I}{V} disease progression}, Journal of
  Pain and Symptom Management 42 (2011), pp. 12--23.

\bibitem{DiBonaventura-et-2012}
M. da{C}osta  {D}i{B}onaventura, S. Gupta, M. Cho, and J. Mrus, \emph{The
  association of {H}{I}{V}\/{A}{I}{D}{S} treatment side effects with health
  status, work productivity, and resource use}, AIDS Care 24 (2012), pp.
  744--755.

\bibitem{dalakas-2009}
M.C. Dalakas, \emph{Toxic and drug-induced myopathies}, Journal of Neurology,
  Neurosurgery \& Psychiatry 80 (2009), pp. 832--838.

\bibitem{dellaportasforster1999}
P. Dellaportas and J.J. Forster, \emph{Markov chain {M}onte {C}arlo model
  determination for hierarchical and graphical log-linear models}, Biometrika
  86 (1999), pp. 615--633.

\bibitem{dellaportastarantola2005}
P. Dellaportas and C. Tarantola, \emph{Model determination for categorical data
  with factor level merging}, Journal of the Royal Statistical Society: Series
  B (Statistical Methodology) 67 (2005), pp. 269--283.

\bibitem{dobra-lenkoski-2011}
A. Dobra and A. Lenkoski, \emph{Copula {G}aussian graphical models and their
  application to modeling functional disability data}, Annals of Applied
  Statistics 5 (2011), pp. 969--993.

\bibitem{dobra-massam-2010}
A. Dobra and H. Massam, \emph{The mode oriented stochastic search
  ({M}{O}{S}{S}) algorithm for log-linear models with conjugate priors},
  Statistical Methodology 7 (2010), pp. 240--253.

\bibitem{dobra-mohammadi-2018}
A. Dobra and A. Mohammadi, \emph{Loglinear model selection and human mobility},
  Annals of Applied Statistics 12 (2018), pp. 815--845.

\bibitem{duracinsky-et-2012}
M. Duracinsky, S. Herrmann, B. Berzins, A. Armstrong, R. Kohli, S.L. Coeur, A.
  Diouf, I. Fournier, M. Schechter, and O. Chassany, \emph{The development of
  {P}{R}{O}{Q}{O}{L}-{H}{I}{V}: An international instrument to assess the
  health-related quality of life of persons living with
  {H}{I}{V}\/{A}{I}{D}{S}}, Journal of Acquired Immune Deficiency Syndromes 59
  (2012), pp. 498--505.

\bibitem{edwardshavranek1985}
D.E. Edwards and T. Havranek, \emph{A fast procedure for model search in
  multidimensional contingency tables}, Biometrika 72 (1985), pp. 339--351.

\bibitem{estanislao-2004}
L. Estanislao, D. Thomas, and D. Simpson, \emph{{H}{I}{V} neuromuscular disease
  and mitochondrial function}, Mitochondrion 4 (2004), pp. 131--139.

\bibitem{fienberg-1970}
S.E. Fienberg, \emph{The analysis of multidimensional contingency tables},
  Ecology 51 (1970), pp. 419--433.

\bibitem{finsterer-mahjoub-2014}
J. Finsterer and S.Z. Mahjoub, \emph{Fatigue in healthy and diseased
  individuals}, American Journal of Hospice and Palliative Medicine 31 (2014),
  pp. 562--575.

\bibitem{fritz-et-2008}
M.S. Fritz and D.P. MacKinnon, \emph{A graphical representation of the mediated
  effect}, Behavior Research Methods 40 (2008"), pp. 55--60.

\bibitem{hidalgo-et-2002}
M. Hidalgo, A. Camozzato, L. Cardoso, C. Preussler, C. Nunes, R. Tavares, M.
  Posser, and M. Chaves, \emph{Evaluation of behavioral states among morning
  and evening active healthy individuals}, Brazilian Journal of Medical and
  Biological Research 35 (2002), pp. 837--842.

\bibitem{hojsgaard-et-2012}
S. Hojsgaard, D. Edwards, and S. Lauritzen, \emph{Graphical Models with {R}},
  Springer-Verlag, New York, 2012.

\bibitem{imai-et-2010}
K. Imai, L. Keele, and D. Tingley, \emph{A general approach to causal mediation
  analysis}, Psychological Methods 15 (2010), pp. 309--334.

\bibitem{imai-et-2013}
K. Imai, D. Tingley, and T. Yamamoto, \emph{Experimental designs for
  identifying causal mechanisms}, Journal of the Royal Statistical Society,
  Series A 176 (2013), pp. 5--51.

\bibitem{jason-choi-2008}
L.A. Jason and C. M., \emph{Dimensions and assessment of fatigue}, in
  \emph{Fatigue Science for Human Health}, Y. Yatanabe, B. Evengard, B.H.
  Natelson, L.A. Jason, and H. Kuratsune, eds., Springer, Tokyo,  2008, pp.
  1--16.

\bibitem{jason-et-2010}
L.A. Jason, M. Evans, M. Brown, and N. Porter, \emph{What is fatigue?
  {P}athological and nonpathological fatigue}, PM\&R 2 (2010), pp. 327--331.

\bibitem{jenkin-et-2006}
P. Jenkin, T. Koch, and D. Kralik, \emph{The experience of fatigue for adults
  living with {H}{I}{V}}, Journal of Clinical Nursing 15 (2006), pp.
  1123--1131.

\bibitem{jong-et-2010}
E. Jong, L. Oudhoff, C. Epskamp, M. Wagener, M. van  Duijn, S. Fischer, and E.
  van  Gorp, \emph{Predictors and treatment strategies of {H}{I}{V}-related
  fatigue in the combined antiretroviral therapy era}, AIDS 24 (2010), pp.
  1387--1405.

\bibitem{jordan-2004}
M.I. Jordan, \emph{Graphical models}, Statistical Science 19 (2004), pp.
  140--155.

\bibitem{keyser-2010}
R.E. Keyser, \emph{Peripheral fatigue: High-energy phosphates and hydrogen
  ions}, PM\&R 2 (2010), pp. 347--358.

\bibitem{koller-friedman-2009}
D. Koller and N. Friedman, \emph{Probabilistic Graphical Models: Principles and
  Techniques}, Adaptive Computation and Machine Learning series, The MIT Press,
  2009.

\bibitem{lauritzen-1996}
S. Lauritzen, \emph{Graphical Models}, Clarendon Press, Oxford, UK, 1996.

\bibitem{leavitt-deluca-2010}
V.M. Leavitt and J. De{L}uca, \emph{Central fatigue: Issues related to
  cognition, mood and behavior, and psychiatric diagnoses}, PM\&R 2 (2010), pp.
  332--337.

\bibitem{leserman-et-2008}
J. Leserman, J. Barroso, B.W. Pence, N. Salahuddin, and J.L. Harmon,
  \emph{Trauma, stressful life events and depression predict {H}{I}{V}-related
  fatigue}, AIDS care 20 (2008), pp. 1258--1265.

\bibitem{mackinnon-et-2007}
D.P. MacKinnon, A.J. Fairchild, and M.S. Fritz, \emph{Mediation analysis},
  Annual Review of Psychology 58 (2007), pp. 593--614.

\bibitem{madiganraftery1994}
D. Madigan and A. Raftery, \emph{Model selection and accounting for model
  uncertainty in graphical models using {O}ccam's window}, Journal of the
  American Statistical Association 89 (1994), pp. 1535--1546.

\bibitem{madiganyork1995}
D. Madigan and J. York, \emph{Bayesian graphical models for discrete data},
  International Statistical Review 63 (1995), pp. 215--232.

\bibitem{madiganyork1997}
D. Madigan and J. York, \emph{Bayesian methods for estimation of the size of a
  closed population}, Biometrika 84 (1997), pp. 19--31.

\bibitem{marconi-et-2013}
V.C. Marconi, B. Wu, J. Hampton, C.E. Ord\'{o}\~{n}ez, B.A. Johnson, D. Singh,
  S. John, M. Gordon, A. Hare, R. Murphy, J. Nachega, D.R. Kuritzkes, C. del
  {R}io, H. Sunpath, and  {South
  ~{A}frica~{R}esistance~{C}ohort~{S}tudy~{T}eam~{G}roup~{A}uthors},
  \emph{Early warning indicators for first-line virologic failure independent
  of adherence measures in a {S}outh {A}frican urban clinic}, {A}{I}{D}{S}
  patient care and {S}{T}{D}s 27 (2013), pp. 657--668.

\bibitem{nixon-et-2005}
S. Nixon, K. O'{B}rien, R. Glazier, and A.M. Tynan, \emph{Aerobic exercise
  interventions for adults living with {H}{I}{V}/{A}{I}{D}{S}}, Cochrane
  Database of Systematic Reviews (2005). John Wiley \& Sons, Ltd.

\bibitem{onen-et-2009}
N.F. \"{O}nen, A. Agbebi, E. Shacham, K.E. Stamm, A.R. \"{O}nen, and E.T.
  Overton, \emph{Frailty among {H}{I}{V}-infected persons in an urban
  outpatient care setting}, Journal of Infection 59 (2009), pp. 346--352.

\bibitem{pearl-2000}
J. Pearl, \emph{Causality: Models, Reasoning, and Inference}, Cambridge
  University Press, New York, 2000.

\bibitem{pence-et-2009}
B.W. Pence, J. Barroso, J.L. Harmon, J. Leserman, N. Salahuddin, and B.G.
  Hammill, \emph{Chronicity and remission of fatigue in patients with
  established {H}{I}{V} infection}, AIDS patient care and STDs 23 (2009), pp.
  239--244.

\bibitem{powers-2017}
S.K. Powers, \emph{Exercise Physiology Theory and Application to Performance},
  6th ed., McGraw Hill, Boston, MA, 2007.

\bibitem{r-manual}
 {R Core Team}, \emph{R: A Language and Environment for Statistical Computing},
  R Foundation for Statistical Computing, Vienna, Austria (2018).
  \urlprefix\url{https://www.R-project.org/}.

\bibitem{salahuddin-et-2009}
N. Salahuddin, J. Barroso, J. Leserman, J.L. Harmon, and B.W. Pence,
  \emph{Daytime sleepiness, nighttime sleep quality, stressful life events, and
  {H}{I}{V}-related fatigue}, The Journal of the Association of Nurses in AIDS
  Care 20 (2009), pp. 6--13.

\bibitem{sliep-et-2001}
Y. Sliep, M. Poggenpoel, and A. Gmeiner, \emph{The experience of {H}{I}{V}
  reactive patients in rural {M}alawi \--- {P}art {I}}, Curationis 24 (2001),
  pp. 56--65.

\bibitem{tarantola2004}
C. Tarantola, \emph{{M}{C}{M}{C} model determination for discrete graphical
  models}, Statistical Modelling 4 (2004), pp. 39--61.

\bibitem{terzian-et-2009}
A.S. Terzian, S. Holman, N. Nathwani, E. Robison, K. Weber, M. Young, R.M.
  Greenblatt, S.J. Gange, and  {Women's Interagency HIV Study}, \emph{Factors
  associated with preclinical disability and frailty among {H}{I}{V}-infected
  and {H}{I}{V}-uninfected women in the era of c{A}{R}{T}}, Journal of Women's
  Health 18 (2009), pp. 1965--1974.

\bibitem{tingley-et-2014}
D. Tingley, T. Yamamoto, K. Hirose, L. Keele, and K. Imai, \emph{mediation: {R}
  package for causal mediation analysis}, Journal of Statistical Software 59
  (2014), pp. 1--38.

\bibitem{voss-et-2007}
J.G. Voss, C.J. Portillo, W.L. Holzemer, and M.J. Dodd, \emph{Symptom cluster
  of fatigue and depression in {H}{I}{V}\/{A}{I}{D}{S}}, Journal of Prevention
  \& Intervention in the Community 33 (2007), pp. 19--34.

\bibitem{wantland-et-2011}
D.J. Wantland, J.P. Mullan, W.L. Holzemer, C.J. Portillo, S. Bakken, and E.M.
  Mc{G}hee, \emph{Additive effects of numbness and muscle aches on fatigue
  occurrence in individuals with {H}{I}{V}/{A}{I}{D} who are taking
  antiretroviral therapy}, Journal of Pain and Symptom Management 41 (2011),
  pp. 469--477.

\bibitem{wantland-et-2008}
D.J. Wantland, W.L. Holzemer, S. Moezzi, S.S. Willard, J. Arudo, K.M. Kirksey,
  C.J. Portillo, I.B. Corless, M.E. Rosa, L.L. Robinson, P.K. Nicholas, M.J.
  Hamilton, E.F. Sefcik, S. Human, M.M. Rivero, M. Maryland, and E. Huang,
  \emph{A randomized controlled trial testing the efficacy of an
  {H}{I}{V}/{A}{I}{D}{S} symptom management manual}, Journal of Pain and
  Symptom Management 36 (2008), pp. 235--246.

\bibitem{whittaker1990}
J. Whittaker, \emph{Graphical Models in Applied Multivariate Statistics}, John
  Wiley $\&$ Sons, 1990.

\bibitem{willard-et-2009}
S. Willard, W.L. Holzemer, D.J. Wantland, Y.P. Cuca, K.M. Kirksey, C.J.
  Portillo, I.B. Corless, M. Rivero-MÃ©ndez, M.E. Rosa, P.K. Nicholas, M.J.
  Hamilton, E. Sefcik, J. Kemppainen, G. Canaval, L. Robinson, S. Moezzi, S.
  Human, J. Arudo, L.S. Eller, E. Bunch, P.J. Dole, C. Coleman, K. Nokes, N.R.
  Reynolds, Y.F. Tsai, M. Maryland, J. Voss, and T. Lindgren, \emph{Does
  ``asymptomatic'' mean without symptoms for those living with {H}{I}{V}
  infection?}, AIDS Care 21 (2009), pp. 322--328.

\bibitem{zhao-et-2010}
X. Zhao, J.G. Lynch, and Q. Chen, \emph{Reconsidering {B}aron and {K}enny:
  Myths and truths about mediation analysis}, Journal of Consumer Research 37
  (2010), pp. 197--206.

\end{thebibliography}

\end{document}